\documentclass[conference]{IEEEtran}
\usepackage{cite}
\usepackage{amsmath,amssymb,amsfonts}
\usepackage{graphicx}

\def\BibTeX{{\rm B\kern-.05em{\sc i\kern-.025em b}\kern-.08em
    T\kern-.1667em\lower.7ex\hbox{E}\kern-.125emX}}

\usepackage{eso-pic}
\AddToShipoutPictureFG*{\AtPageUpperLeft{\put(36,-20){\parbox[t]{540pt}{\fontsize{8}{9}\selectfont
\textcopyright\ 2026 IEEE. Permission from IEEE must be obtained for all other uses, in any current or future media, including reprinting/republishing this material for advertising or promotional purposes, creating new collective works, for resale or redistribution to servers or lists, or reuse of any copyrighted component of this work in other works.\\
Author accepted manuscript. Accepted for publication in the 2026 IEEE Biomedical Circuits and Systems Conference (BioCAS).
}}}}

\begin{document}

\title{\vspace{14pt}NORA: Neuro-headstage for Online Recording and Actuation in Freely Behaving Animal Models}

\author{\IEEEauthorblockN
        {Savanna~Blade\IEEEauthorrefmark{1},
        Yuhan~Hou\IEEEauthorrefmark{1},
        He~Jia\IEEEauthorrefmark{1},
        Laura Kondrataviciute\IEEEauthorrefmark{2},
        Runtian~Yang\IEEEauthorrefmark{1},\\
        Suneil K. Kalia\IEEEauthorrefmark{2}\IEEEauthorrefmark{3},
        and~Xilin~Liu\IEEEauthorrefmark{1}
        }
        
\IEEEauthorblockA{
\IEEEauthorrefmark{1}Department of Electrical and Computer Engineering (ECE), 
\IEEEauthorrefmark{2}Department of Surgery, University of Toronto, \\Toronto, ON, Canada
\IEEEauthorrefmark{3}Krembil Research Institute, Toronto Western Hospital, Toronto, ON, Canada
}
}

\maketitle

\begin{abstract} 

This paper presents NORA, a compact wireless neuro-headstage designed for simultaneous neural recording and stimulation in freely behaving animal models. Separate recording and stimulation interfaces, each managed with a dedicated MCU and wireless module, are integrated into a fully head-mounted device compatible with standard Omnetics connectors; this separation allows for simultaneous recording and stimulation without sample-dropping or stimulation waveform distortion. To minimize stimulation artifacts and cross-domain interference, the recording and stimulation subsystems are electrically isolated with independent power domains and dedicated wireless communication paths. NORA has been validated through \textit{in vivo} animal experiments, demonstrating acquisition of meaningful neural features with results comparable to a commercial wireless headstage. Together, these features establish a practical platform for artifact-reduced neuromodulation research. 

\end{abstract}


\section{Introduction}

Bidirectional neural recording and stimulation technologies are essential tools for studying brain function and developing neuromodulation therapies. To capture neural activity under naturalistic behavioral conditions, it is necessary to develop neural interfaces that allow animals to roam freely without restriction by tethers~\cite{Yang2021, Newman2025}. Recent advances have enabled multichannel wireless neural acquisition and stimulation in rodent models~\cite{Pederson_2019,Ouyang2023,Hoang_Kang_Koh_Park_2025} and miniature wireless platforms for closed-loop modulation~\cite{Lee2025AuditorySleep}. More broadly, multimodal wireless sensing systems have enabled home-based monitoring of rehabilitation exercises~\cite{Yang2024Rehabilitation}. Despite this progress, obstacles remain that impede the experimental utility of existing wireless neural interfaces. In miniaturized systems, shared power rails and communication modules can introduce electrical coupling and stimulation artifacts that degrade signal quality~\cite{Zhou_Santacruz_Johnson_Alexandrov_Moin_Burghardt_Rabaey_Carmena_Muller_2019}; in embedded interfaces, shared processing resources can cause timing conflicts that result in charge-imbalanced stimulation, damaging brain tissue~\cite{Merrill_Bikson_Jefferys_2005}. Integration compatibility further limits adoption, as many wireless platforms require custom probe interfaces or specialized infrastructure~\cite{Ouyang2023, Lee_Koripalli_Jia_Acosta_Sendi_Choi_Ghovanloo_2018} that is incompatible with established experimental workflows.

\begin{figure}[!t]
    \centering
    \includegraphics[width=\linewidth]{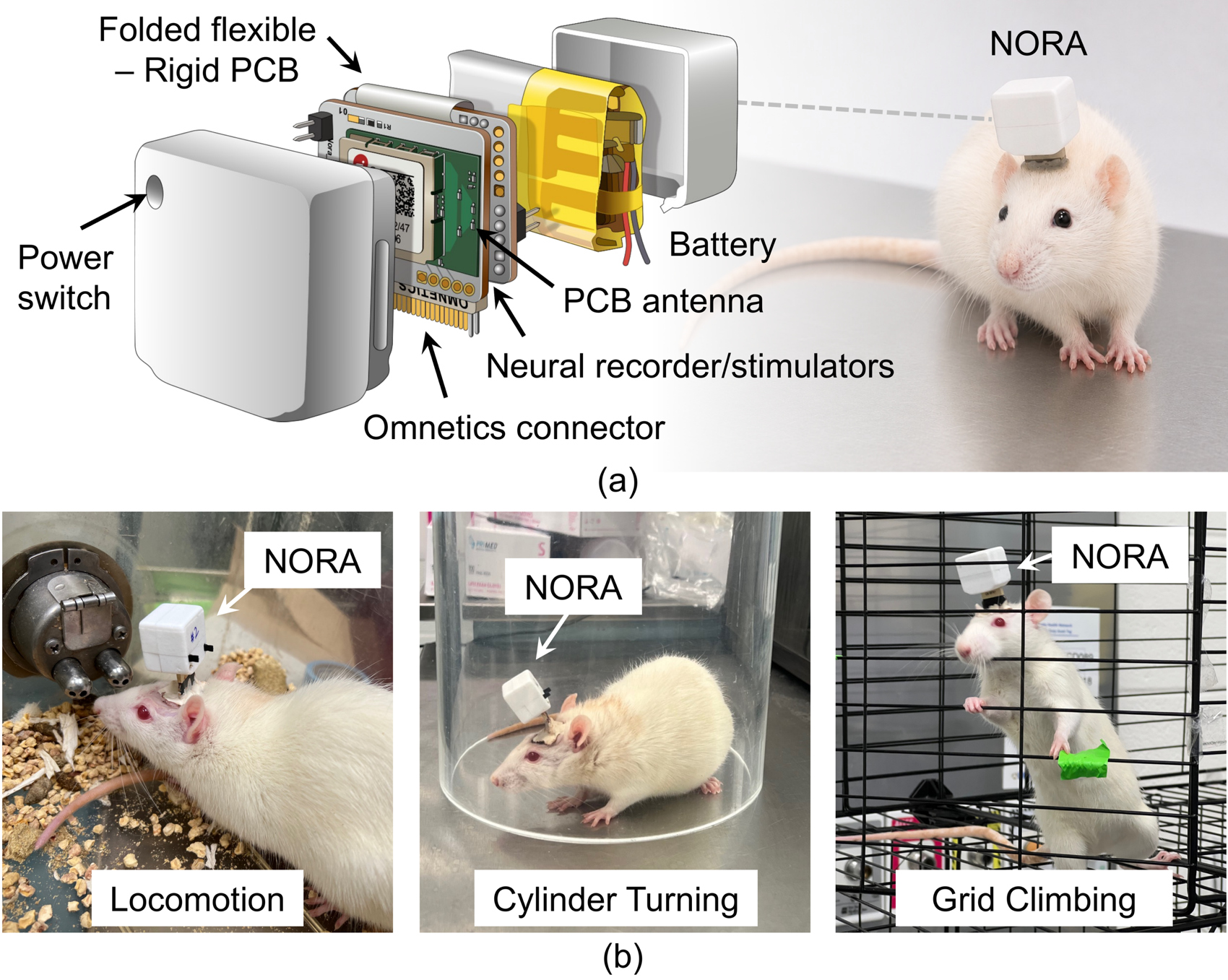} \vspace{-7mm}
    \caption{(a) Illustration of NORA’s design and construction. (b) Representative photographs showing wireless operation of NORA during diverse behavioral paradigms in freely behaving animals.}
    \label{fig:intro}
\end{figure}

This paper presents NORA, a wireless \textbf{\underline{N}}euro-headstage for \textbf{\underline{O}}nline 
\textbf{\underline{R}}ecording and \textbf{\underline{A}}ctuation in freely behaving animal models.  
A high-level illustration of NORA is shown in Fig.~\ref{fig:intro}. The system integrates two off-the-shelf neural interface chips (RHD2216 and RHS2116, Intan Technologies), with five channels of each used for recording and stimulation, respectively. Each chip is controlled by a dedicated wireless microcontroller (MCU) (nRF5340, Nordic Semiconductor), eliminating timing conflicts between acquisition and stimulation workloads. The recording and stimulation modules operate from separate power supplies to minimize electrical coupling and stimulation artifacts. To improve experimental integration, the system features a 36-pin Omnetics interface compatible with a wide variety of commercial neural probes.

The primary contributions of this work are the following: (1) a compact, fully head-mounted wireless neuro-headstage for simultaneous neural recording and stimulation in freely behaving animal models; (2) a duplex wireless architecture with isolated recording and stimulation domains for artifact-reduced bidirectional neural interfacing; (3) compatibility with standard Omnetics-based neural probes to support experimental integration; and (4) \textit{in vivo} validation demonstrating reliable acquisition of neural features, including single-unit spiking activity and pathological power signatures. 

\section{System Architecture}

\begin{figure}[!ht]
    \centering
    \includegraphics[width=\linewidth]{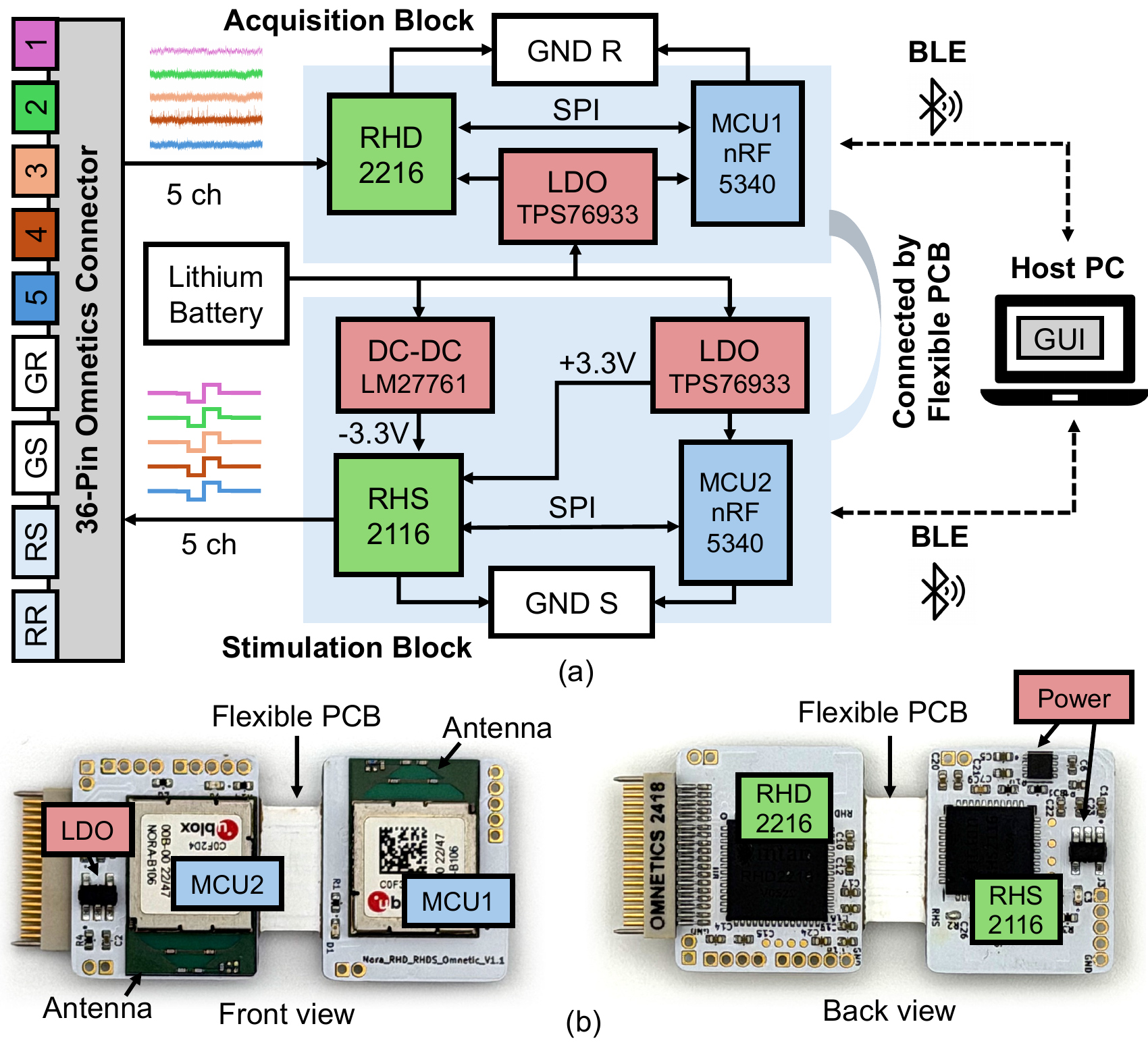}
    \caption{(a) Block diagram of the hardware. (b) Front and back views of the assembled rigid-flex PCB with key blocks highlighted.}
    \label{fig:architecture}
\end{figure}

Fig.~\ref{fig:architecture} provides a general overview of the platform. The system comprises two main functional blocks: an acquisition block and a stimulation block. Each of the two functional blocks is supplied with its own power and ground rails, provided by a single 60\,mAh LiPo battery. To minimize the overall form factor, the system is packaged in a compact 3D-printed case and a folded rigid-flex PCB composed of two 15\,mm\,x\,18\,mm boards, as shown in Fig.~\ref{fig:intro}. The folded assembly stacks the neural recorder/stimulator circuitry, MCU, Omnetics connector, and battery within the case, weighing 8.22\,g altogether. The 36-pin Omnetics connector features 5 pins that can be interchangeably configured as either recording or stimulation channels, with separate reference and ground pins for the RHD2216 and RHS2116. Two nRF5340 MCUs coordinate data acquisition from the RHD2216 and stimulation delivery by the RHS2116. Each nRF5340 communicates with a graphical user interface (GUI) application running on a host PC via independent BLE connections. 

The software framework, including the nRF5340 firmware and GUI, is designed to support upstream data transfer as well as downstream stimulation and configuration commands for two or more separate BLE connections. This is accomplished via a custom data transfer protocol, handled on each side of the transfer by an abstract ``manager'' layer responsible for coordinating the reception and transmission of data. Fig.~\ref{fig:firmware} contains a block diagram of the firmware and illustrates the interactions of these different components of the communication protocol. A further description of the two ``manager'' layers in the communication pathway is provided:

\begin{figure}[!ht]
    \centering
    \includegraphics[width=\linewidth]{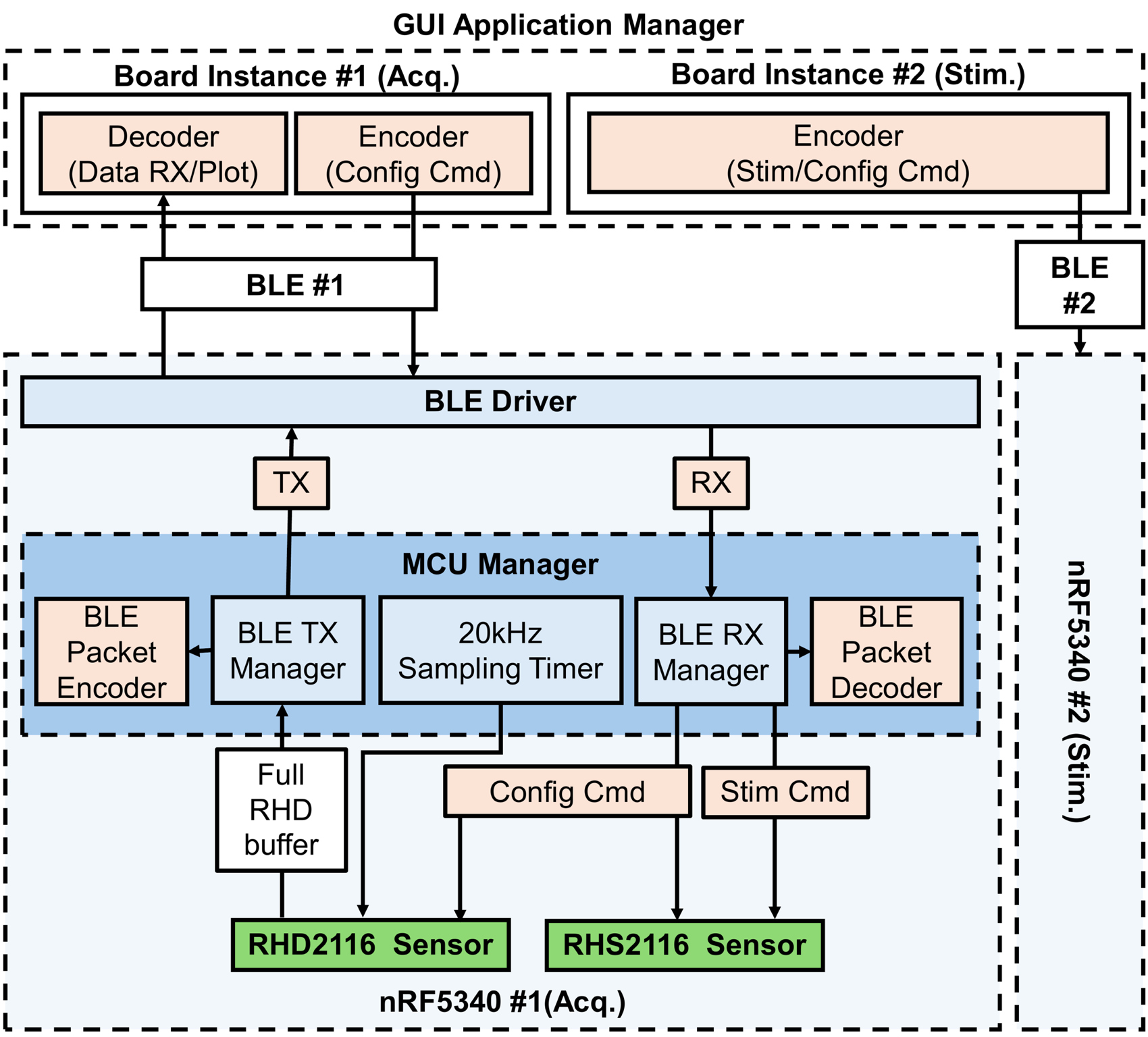} \vspace{-5mm}
    \caption{Prototypical firmware block diagram and illustration of communication between the GUI and two nRF5340 MCUs. General firmware architecture is identical on both nRF5340 MCUs.} \vspace{-3mm}
    \label{fig:firmware}
\end{figure}

    \subsubsection{\textbf{MCU Manager}} The MCU manager layer in the firmware sits between the BLE and sensor drivers. The operations of the firmware are detailed as follows:
        (1) Initialization: On power-up, the nRF5340 Network core begins advertising over BLE with a custom connection string specific to each nRF5340. The MCU manager is initialized and a 20\,kHz sampling hardware timer is enabled.
        (2) Sampling and TX: Every 50\,$\mu$s, a timer interrupt triggers a sampling callback. The sampling callback in the RHD2216 sensor driver sends a command over SPI to the RHD2216 on every trigger; either a NOOP command or a sampling command, based on the set sampling rate. As a result, the maximum sampling rate for the RHD2216 is 20\,kHz divided by the number of enabled channels. The RHD2216 sensor driver handles sample accumulation in a 240-byte buffer that, when full, is pushed to the MCU manager. The MCU manager constructs a BLE packet with the payload and a custom packet header; the packet is then added to a FIFO queue, which is consumed as data is transmitted in an independent BLE thread.
        (3) RX: Upon BLE packet RX, the MCU manager decodes the custom header, triggering the appropriate command callback. When a stimulation command is received, a stimulation hardware timer is initialized in the RHS2116 sensor driver that uses compare channel interrupts to manage stimulation timing.
    Even with stimulation timer interrupts assigned higher priority than the 20\,kHz sampling timer, enabling both simultaneously can introduce timing conflicts that result in unbalanced stimulation waveforms. Distributing acquisition and stimulation responsibilities across two dedicated MCUs eliminates these conflicts, ensuring precise and reliable stimulation pulse timing.
    \subsubsection{\textbf{GUI Application Manager}} The application manager is the central backend data structure in the GUI. Every nRF5340 device that is connected to the GUI belongs to the application manager as a ``board instance'' with its own BLE connection. When a configuration or stimulation command is input by the user, the board instance manager encodes the command with the appropriate custom header and handles BLE TX. On the reception of a data packet, the board instance manager handles the decoding and visualization of the incoming data stream.
There are several advantages offered by this modular framework. Both MCUs can run the same firmware with minimal code changes (save for modifying the BLE communication string), allowing for straightforward modification of common operations. This framework also allows control of both MCUs from the GUI using an intuitive and repeatable workflow. 

\section{Experimental Results}
\begin{figure}[!ht]
    \centering
    \includegraphics[width=\linewidth]{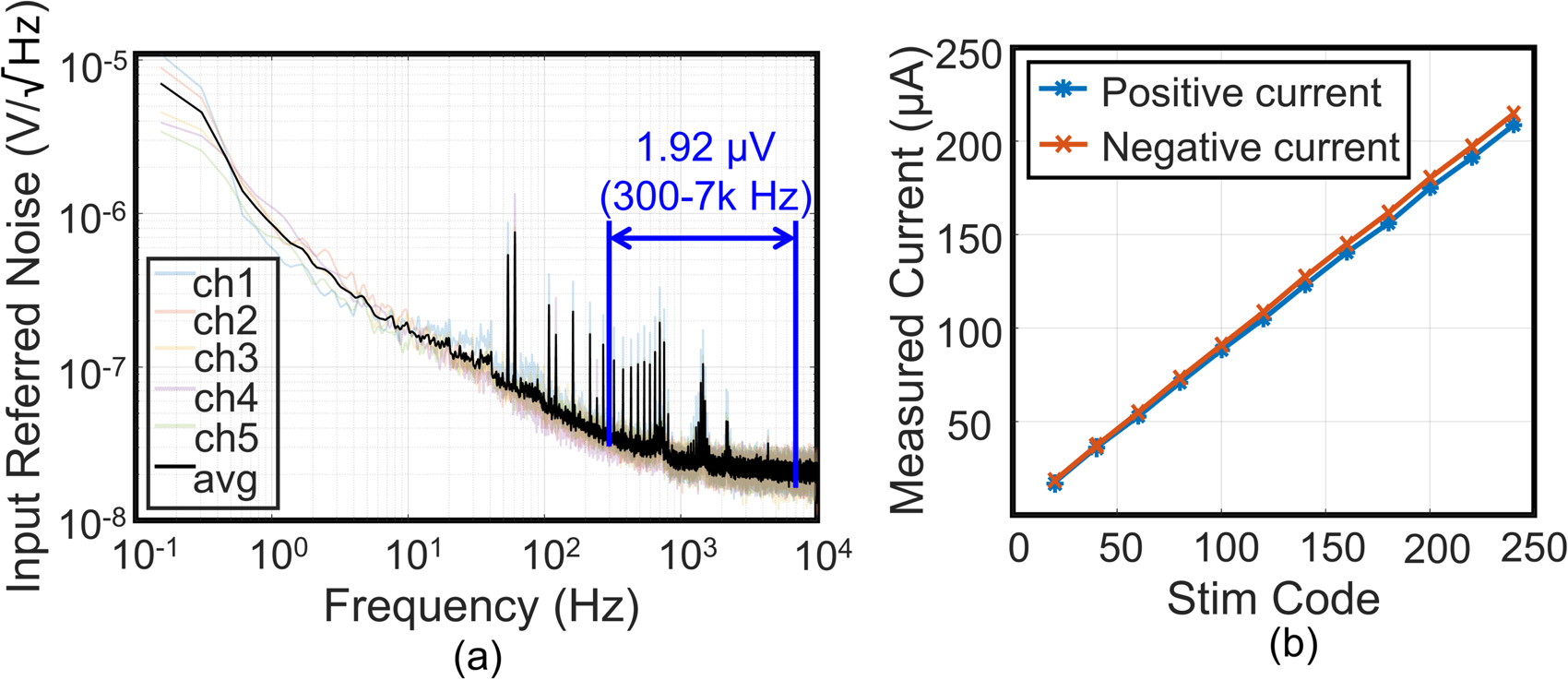} \vspace{-5mm}
    \caption{NORA headstage characterization. (a) Input-referred noise PSD measured from five recording channels through Bluetooth readout, with the averaged PSD shown in black. (b) Measured stimulation current versus programmed current for positive and negative phases.}
    \label{fig:device_meas}
\end{figure}

The electrical performance of NORA was first characterized on the bench. Fig.~\ref{fig:device_meas}(a) shows the input-referred noise power spectral density (PSD) measured from five recording channels, with data transmitted via Bluetooth. The averaged input-referred noise across the five channels is 1.92\,$\mu$V over 300\,Hz–7\,kHz, the neural spike acquisition band. Stimulation output was characterized by sweeping the programmed current settings and measuring the delivered biphasic pulse current. As shown in Fig.~\ref{fig:device_meas}(b), the measured positive and negative pulse currents closely follow the programmed values; calibration can be readily applied to achieve the desired charge balance.

To validate the NORA headstage and demonstrate its acquisition and stimulation capabilities, we performed several \textit{in vivo} experiments in rat models. Experiments were conducted under Canadian Council on Animal Care AUP 3818.

\begin{figure}[!ht]
    \centering
    \includegraphics[width=\linewidth]{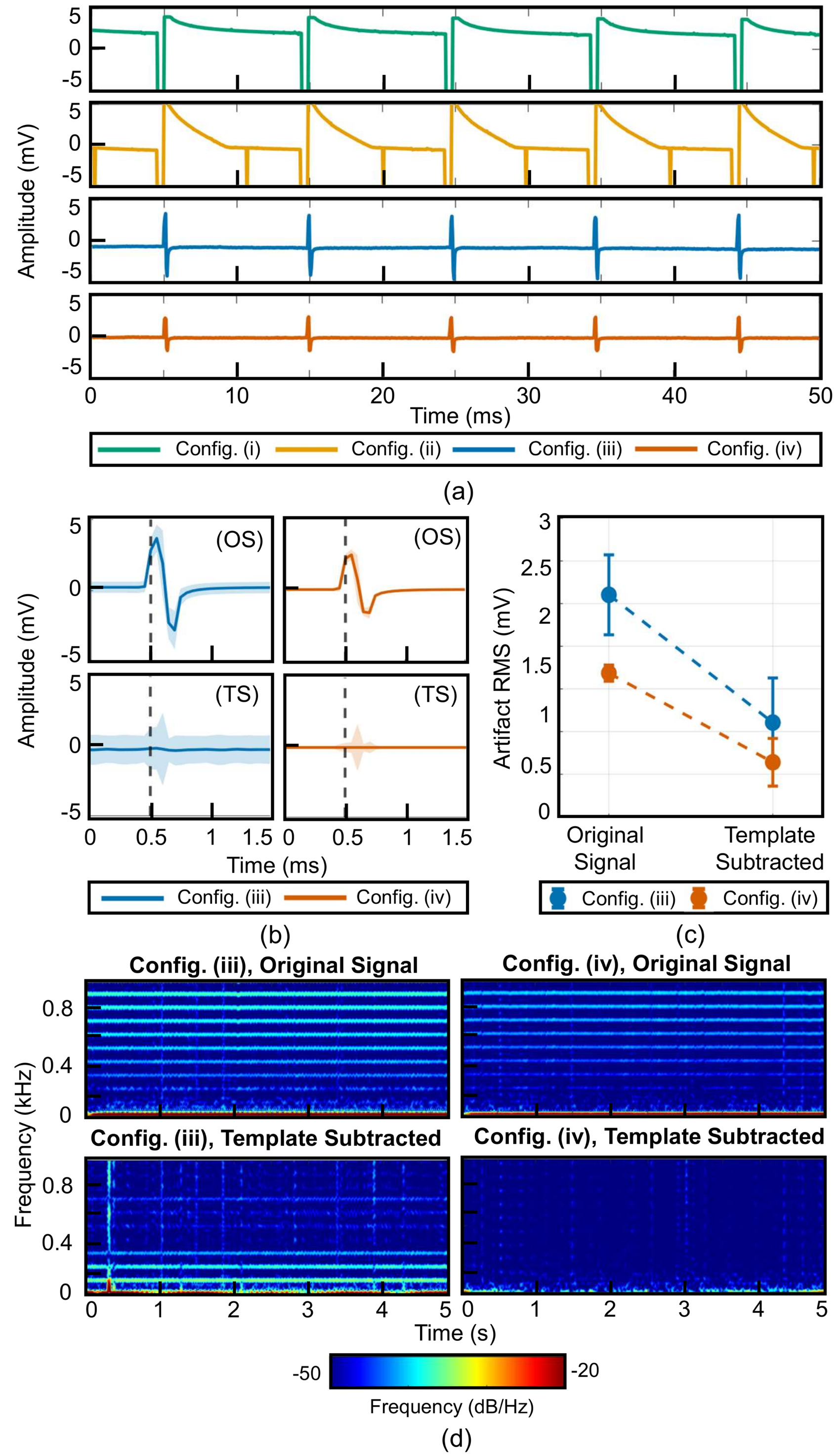} \vspace{-8mm}
    \caption{(a) Stimulation artifact comparison for different recording/stimulation grounding configs measured \textit{in-vivo} (n=1). (b) Comparison of average and std. artifact waveform for configs. (iii) and (iv), before (OS, top row) and after (TS, bottom row) template subtraction. 996 artifact epochs used per configuration for analysis. (c) RMS amplitude of artifact epochs before and after template subtraction for configs. (iii) and (iv). (d) Spectrogram of 5-second stimulation period, pre and post template subtraction for configs. (iii) and (iv).}
    \label{fig:comparison}
\end{figure}

The purpose of the first experiment was to validate two design choices: first, the decision to use the RHD2216 for acquisition and RHS2116 solely for stimulation, even though the RHS2116 is capable of both recording and stimulation; second, separating the acquisition and stimulation grounds. Recordings were taken from an adult female Sprague-Dawley rat under 1.5–2\,\% isoflurane anesthesia. Balanced biphasic stimulation was applied at a frequency of 100\,Hz with an amplitude of 100\,$\mu$A. Fig.~\ref{fig:comparison}(a) provides the results from applying stimulation using four separate recording configurations: (i) using the RHS2116 for both recording and stimulation without any artifact recovery measures; (ii) using the single RHS2116 for recording and stimulation with low-frequency cutoff shifting implemented, where the RHS analog filter high-pass cutoff is raised from 0.1\,Hz to 1\,kHz for 5\,ms after stimulation; (iii) using the RHD2216 for acquisition and RHS2116 for stimulation with a shared ground, and (iv) using the two with separate grounds. In configuration (i), the artifact persists throughout the inter-stimulus interval. Low-frequency cutoff shifting reduces recovery time slightly, but artifact amplitude remains large; additionally, a secondary transient is introduced when the cutoff returns to its recording value, visible as an abrupt deflection following the recovery period. For the dual RHD2216 and RHS2116 configurations (iii) and (iv), the average artifact waveform amplitudes resulting from the shared and separate ground configurations were 7.62 $\pm$ 0.76\,mV$_{\mathrm{pp}}$ and 4.30 $\pm$ 0.16\,mV$_{\mathrm{pp}}$, respectively (Fig.~\ref{fig:comparison}(b)). The difference in average artifact amplitude between configurations is consequential for post-hoc artifact removal: larger artifacts introduce greater trial-to-trial waveform variability, meaning template subtraction leaves a larger residual. This is reflected in the residual RMS comparison (Fig.~\ref{fig:comparison}(c)), where the shared-ground configuration shows considerably higher residual contamination (1.05 $\pm$ 0.45\,mV for (iii), 0.64 $\pm$ 0.28\,mV for (iv)) despite identical subtraction procedures. Fig.~\ref{fig:comparison}(d) shows the frequency-domain effect of template subtraction for both configurations. Configuration (iii) shows reduced efficacy at suppressing stimulation harmonics.

\begin{figure}[!ht]
    \centering
    \includegraphics[width=\linewidth]{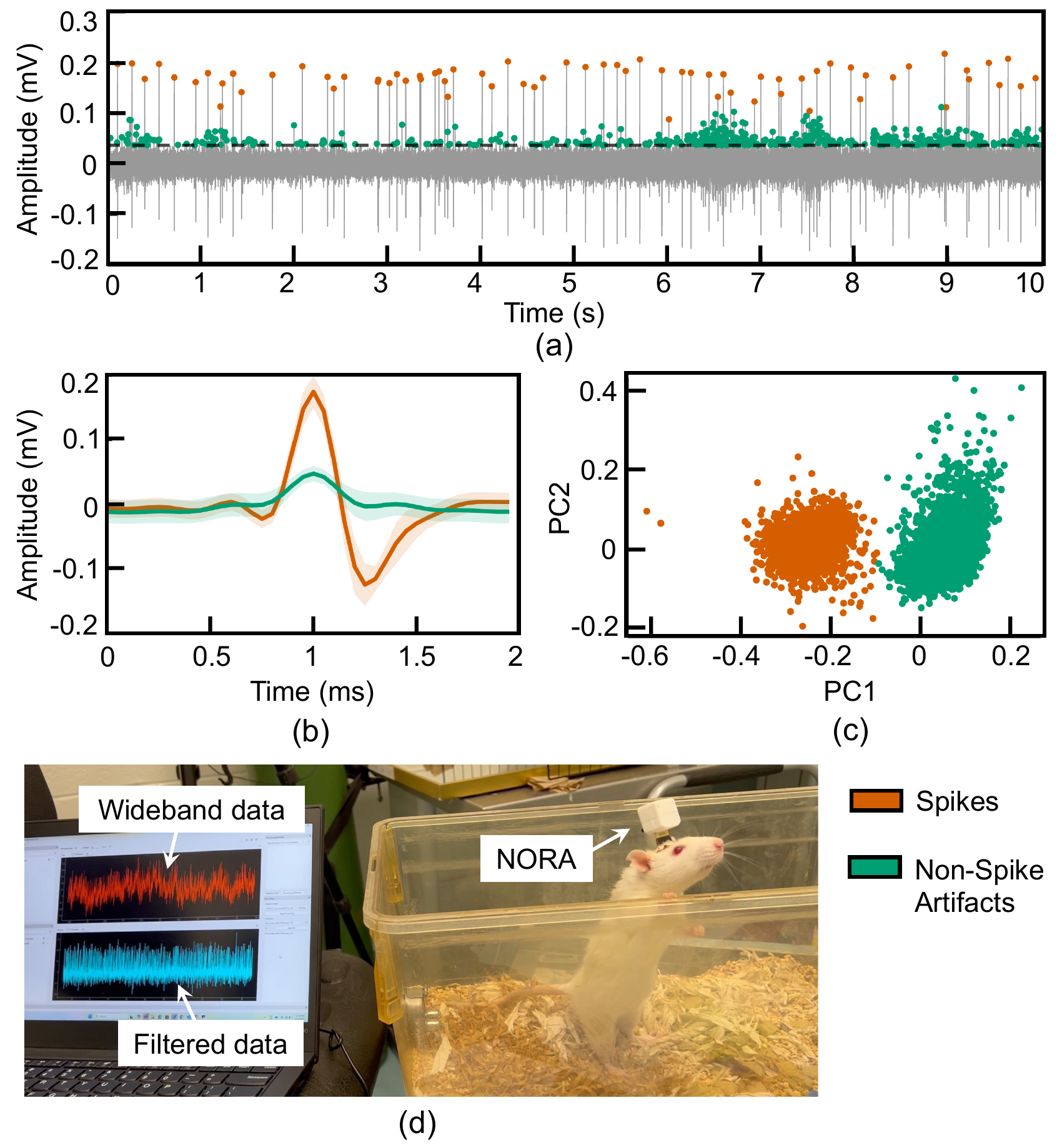} \vspace{-7mm}
    \caption{(a) Spike occurrences in data from one A53T rat (250-3000\,Hz bandpass filtered); (b) Average and std. waveform of spike and non-spike artifacts, using 1026 identified spike epochs and 2903 artifact epochs; (c) PCA results (silhouette score of 0.77); (d) A photo showing real-time wireless spike recording and plotting in a freely behaving rat.}
    \label{fig:spiking}
\end{figure}

The second experiment served to demonstrate the acquisition reliability of the NORA system. Recordings were taken from six awake and freely moving adult female Sprague-Dawley rats with recording sites distributed within the subthalamic nucleus (STN). Four rats were administered a unilateral injection of an adeno-associated virus (AAV) encoding human mutant A53T alpha-synuclein into the substantia nigra to produce electrophysiological signatures of Parkinson's disease (PD), including pathological oscillations in the beta range (13-30\,Hz); the remaining two rats received an empty vector (EV), acting as controls~\cite{Kondrataviciute2026}. To verify that the NORA device is capable of capturing single-neuron activity, recordings were taken at the NORA's maximum sampling rate (20\,kHz). Principal component analysis (PCA) was performed to separate spikes from non-spike artifacts. Fig.~\ref{fig:spiking} presents 8.0\,Hz spiking activity captured in a single-channel recording from an A53T $\alpha$-Syn-overexpressing animal, with a signal-to-noise ratio of 2.1 and 0.0\% inter-spike interval violations.

\begin{figure}[!ht]
    \centering
    \includegraphics[width=\linewidth]{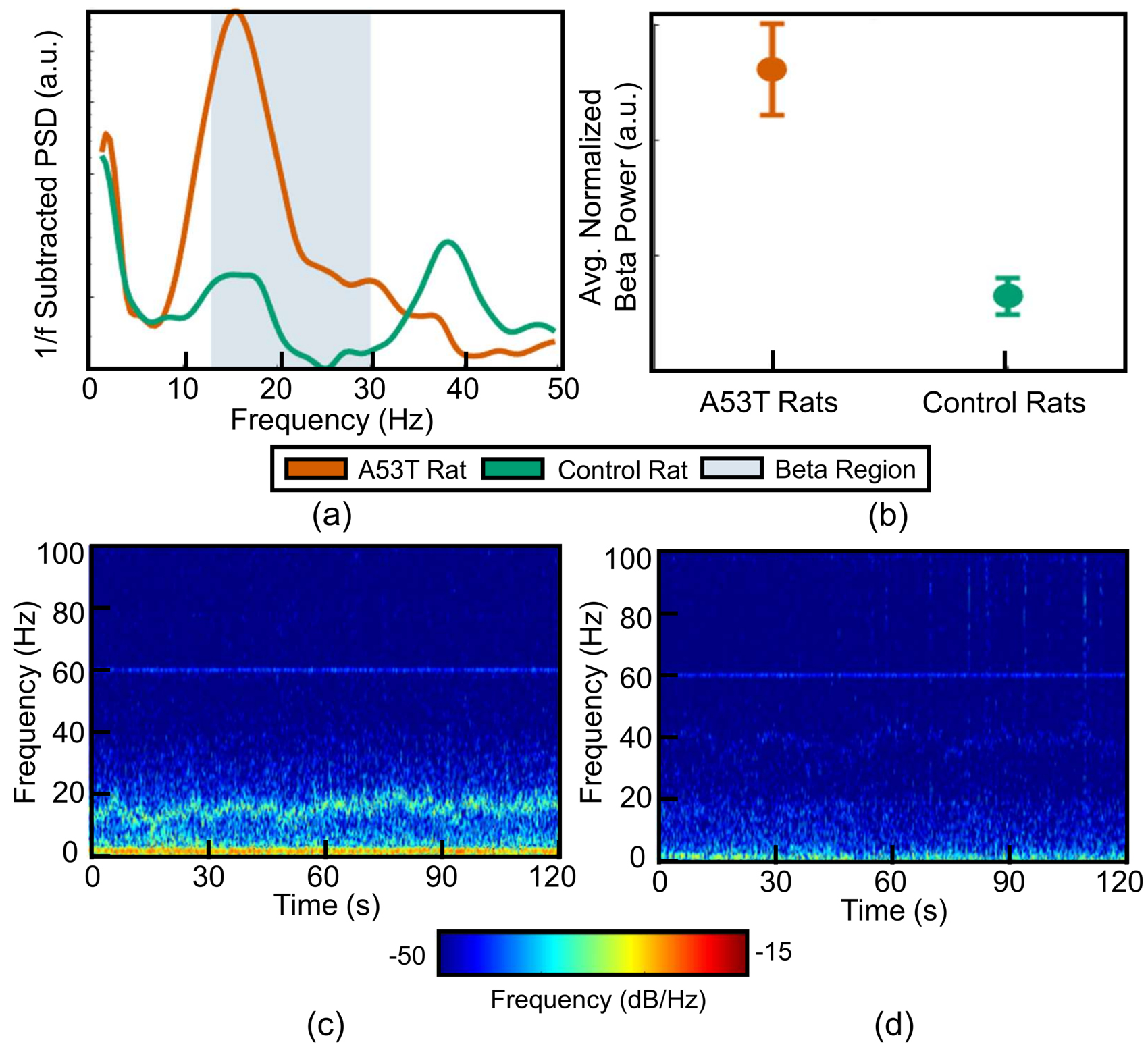} \vspace{-8mm}
    \caption{(a) Representative 1/f-subtracted PSD of A53T and control rat LFPs. Beta region (13\,Hz-30\,Hz) is highlighted. (b) Average normalized beta power comparison across all A53T rats (n=4) and control rats (n=2). (c) Representative A53T rat spectrogram. (d) Representative control rat spectrogram.}
    \label{fig:exp_results}
\end{figure}

A spectral comparison was performed to evaluate the NORA's ability to capture the elevated beta power observed in local field potentials (LFPs) of freely behaving A53T $\alpha$-Syn-overexpressing rats six weeks post-AAV injection, as previously reported using the commercial NeuLive wireless acquisition system~\cite{Kondrataviciute2026}. Following sampling rate calibration to align power line noise to 60\,Hz, recordings from the A53T group showed average beta power elevation consistent with prior findings (Fig.~\ref{fig:exp_results}), confirming that the NORA captures signal features comparable to a commercial headstage.

\section{Conclusion}
NORA is a compact, wireless neural interface supporting up to 5 bidirectional channels and an aggregate 20\,kHz sampling rate, featuring a power-rail separation strategy to minimize stimulation artifacts. It integrates seamlessly into existing experimental workflows through a standard 36-pin Omnetics connector. NORA's functionality has been validated in anesthetized and freely moving rodent models, confirming its artifact mitigation and acquisition capabilities.

The design of the custom communication platform between the GUI and the nRF5340 firmware is highly extensible beyond the requirements of the reported system. While each nRF5340 in this system only interfaces with one RHD/RHS device, the firmware design enables the coordination of multiple sensors sampling at different rates, and/or multiple stimulation sources, all from a single MCU. Future directions include leveraging the platform to develop closed-loop devices with diverse sensing and stimulation modalities. 

\section*{Acknowledgment}

We thank Xiao Wu, Yinfei Li, and Xinyu Song for their technical contributions and assistance in testing. 

\bibliographystyle{IEEEtran}
\bibliography{ref}

\end{document}